# Simultaneous control of the Dzyaloshinskii-Moriya interaction and magnetic anisotropy in nanomagnetic trilayers

A.L. Balk[1,2,3], K-W. Kim[1,2], D.T. Pierce[1], M.D. Stiles[1], J. Unguris[1], S.M. Stavis[1,*]

[1]Center for Nanoscale Science and Technology, National Institute of Standards and Technology, Gaithersburg, MD 20899 USA, [2]Maryland NanoCenter, University of Maryland, College Park, MD 20742 USA, [3]National High Magnetic Field Laboratory, Los Alamos National Laboratory, Los Alamos, NM 87545 USA, *Address correspondence to sstavis@nist.gov

**Magneto-optical Kerr effect (MOKE) microscopy measurements of magnetic bubble domains demonstrate that Ar[+] irradiation around 100 eV can tune the Dzyaloshinskii-Moriya interaction (DMI) in Pt/Co/Pt trilayers. Varying the irradiation energy and dose changes the DMI sign and magnitude separately from the magnetic anisotropy, allowing tuning of the DMI while holding the coercive field constant. This simultaneous control emphasizes the different physical origins of these effects. To accurately measure the DMI, we propose and apply a physical model for a poorly understood peak in domain wall velocity at zero in-plane field. The ability to tune the DMI with the spatial resolution of the Ar[+] irradiation enables new fundamental investigations and technological applications of chiral nanomagnetics.**

The Dzyaloshinskii-Moriya interaction (DMI) is a chiral energy term leading to interesting nanoscale magnetic phenomena [1-4]. Examples include skyrmions [5-10], domain wall motion [11-15], helical spin textures [16,17], and spin-orbit torque magnetization switching [18-20]. However, such phenomena depend on a combination of the DMI and non-chiral magnetic effects such as anisotropy [21-23], Zeeman energy [4,24,25] and dipolar interaction [26-28]. While previous studies have demonstrated several methods for controlling the DMI [29-32], these also lead to variations in magnetic anisotropy and exchange interaction, making it difficult to reach a specific location in the phase space of micromagnetic energy parameters. Furthermore, these methods cannot be applied to different regions of the same sample, requiring separate samples to vary the DMI. Addressing these limitations, we demonstrate post-growth Ar[+] irradiation around 100 eV as a method to tune the sign and magnitude of the DMI, separately from the coercive field $\mu_0 H_c$, in a single trilayer of Pt/Co/Pt, a model nanomagnetic system with perpendicular interfacial anisotropy.

First, we report the sample growth and Ar[+] irradiation procedures that we use to tune the properties of the trilayer. Then, we describe the magneto-optical Kerr effect (MOKE) microscopy technique that we use to measure domain wall motion and infer the effective DMI field $\mu_0 H_{DMI}$ in the trilayer. Some of our data sets show a previously observed but unexplained peak in domain wall velocity at zero in-plane field, for which we propose a physical model. We fit our data with this model and demonstrate that we can simultaneously control $\mu_0 H_{DMI}$ and $\mu_0 H_C$ by varying the energy and dose of Ar[+] irradiation, enabling us to implement arbitrary combinations of these magnetic parameters. Last, we attribute the separate tuning of $\mu_0 H_{DMI}$ and $\mu_0 H_C$ [33-36] to the distinct effects of etching the top Pt layer and modifying the disorder of the Co/Pt interfaces by Ar[+] irradiation.

We study a single trilayer of Pt(35 nm)/Co(0.8 nm)/Pt(1.7 nm) that we sputter on a p-type silicon wafer (Supplemental Material S1). We use a shadow mask to irradiate the trilayer with a spatially varying dose of Ar[+] at a range of energies $E_{Ar+}$ from 50 eV to 140 eV in increments of 5 eV, and then remove the sample from vacuum for measurement in air. The irradiation reduces the as-grown value of $\mu_0 H_C \approx 80$ mT to values ranging to 0 mT where spontaneous domain fluctuations [37] and a spin reorientation transition [35] occur. For this trilayer and these energies, the spin reorientation transition occurs with Ar$^+$ doses of approximately 2 × 10$^{15}$ cm$^{-2}$ to 2 × 10$^{16}$ cm$^{-2}$.

We measure the DMI of different regions of this trilayer using MOKE microscopy of the expansion of magnetic bubble domains with applied out-of-plane and in-plane fields $B_z$ and $B_y$. In this technique, $B_z$ creates and expands bubbles while the DMI and $B_y$ lead to asymmetry of this expansion (Fig. 1(a)) by modifying the energy of the Néel walls surrounding the bubbles [29,38,39]. For example, the left side of the left bubble in Fig. 1(a) expands faster in the direction of $B_y$ because this Néel wall has a lower energy due to $B_y$, corresponding to $\mu_0H_{DMI}$ > 0 mT. Building on the earlier reports, we pulse $B_z$ to generate and annihilate magnetic bubbles at a repetition frequency of 50 Hz to 150 Hz, which is higher than the imaging frequency of 10 Hz. Therefore, each MOKE micrograph (Fig. 1, (b,c)) shows a bubble at its maximum size, from the average of 5 to 15 expansions. While applying pulses of $B_z$, we apply $B_y$ as a triangle wave with a frequency of 50 mHz, which is much slower than the $B_z$ repetition frequency. Therefore, the bubbles that we measure accurately reflect how $B_y$ affects their expansion. Supplemental Material S2 presents details of the field excitation technique. To measure domain wall displacements in real time, we take a cross-sectional strip of each bubble (Fig. 1(b,c), blue and red rectangles), average these strips across the $x$ direction, (Fig. 1(d,e), blue and red markers), and fit each resulting profile to arctangent functions (Fig. 1(d,e), black lines) which empirically model the optically broadened profile of the domain walls (Supplemental Material S3) and superresolve their positions in real time. We divide the domain wall displacements by the duration of the $B_z$ pulses to obtain the mean velocity $v$ of each domain wall during bubble expansion.

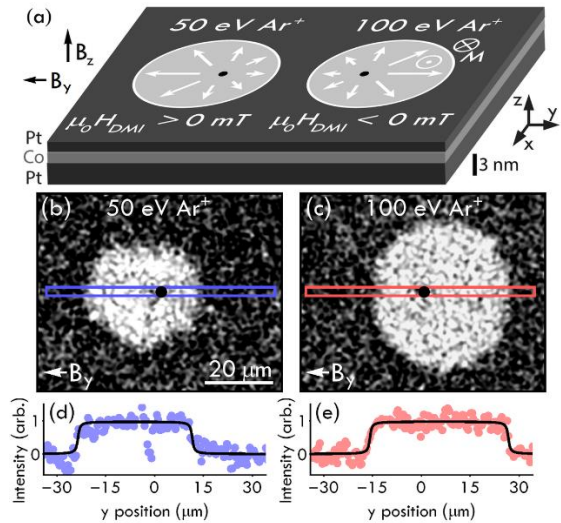

Fig. 1: Ar$^+$ irradiation of a Pt/Co/Pt trilayer controls the sign of the Dzyaloshinskii-Moriya interaction (DMI). (a) A simplified schematic shows magnetic bubble domains with positive $M_z$ (light gray circles) in trilayer regions irradiated with 50 eV Ar$^+$ or 100 eV Ar$^+$, expanding asymmetrically (white arrows) in opposite $y$ directions from nucleation sites (black dots) in response to applied fields $B_z$ and $B_y$. The direction of this asymmetric expansion indicates the sign of $\mu_0H_{DMI}$. (b,c) Magneto-optical Kerr effect (MOKE) micrographs showing corresponding experimental results. We have spatially filtered and removed backgrounds from these representative micrographs for clarity. The Ar$^+$ dose is approximately 2 × 10$^{16}$ cm$^{-2}$ for (b) and approximately 2 × 10$^{15}$ cm$^{-2}$ for (c). Black dots indicate the approximate position of bubble nucleation. Blue and red rectangles indicate regions from which we take profiles to determine domain wall displacements. (d,e) Blue and red markers are profiles from boxed regions in (b) and (c)

averaged across the *x* direction. Black curves are arctangent fits used to extract domain wall displacements with standard uncertainties of approximately 400 nm.

To perform a measurement of the DMI, we measure *v* as a function of $B_y$. The domain wall comprising the bubble has opposite symmetry on either side of the bubble with respect to $B_y$, so that in the reference frame of the domain wall the effective magnetic field $B_{y(eff)}$ is opposite in sign for each side of the bubble. (Fig. 2, inset). Therefore, we resample the raw data, which includes multiple cycles of $B_y$, by averaging data points with $B_{y(eff)}$ within 1 mT of each other (Fig. 2). This data comes from a region of the trilayer that we expose to 125 eV Ar$^+$, with $\mu_0 H_c \approx 3$ mT, and is an average of 12 $B_y$ cycles. The similar behavior of the domain wall on both sides of the bubble indicates the preservation of the domain wall chirality around the bubble, which is typical for all bubbles in this study. The data does not show evidence for chiral damping [40], as it does not have a significant asymmetry about its minimum value (Supplemental Material S4).

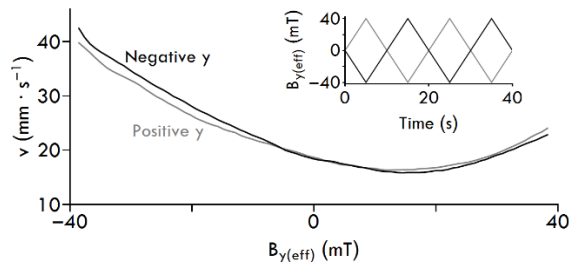

Fig. 2: Representative data shows a simultaneous measurement of the velocity *v* of a domain wall on the positive *y* (gray) and negative *y* (black) sides of a bubble. The domain wall moves similarly on both sides of the bubble in response to the effective magnetic field $B_{y(eff)}$, with a reversed sign for opposite sides. We attribute the small difference between the curves to a misalignment of $B_y$ which adds a *z* component to the total field. Standard uncertainties of *v* are approximately 0.2 mm·s$^{-1}$. Inset: Simulated data shows that $B_{y(eff)}$ is opposite in sign for the domain wall on each side of the bubble, due to the opposite spatial symmetry.

For a systematic study of the DMI, we obtain over 400 *v* curves such as those in Fig. 2, although averaging over only two cycles of $B_y$. To compare the results from many bubbles which may not have the same *v*, we normalize each curve to its value at $B_{y(eff)} = 0$, $v_{norm} = v/v_{(By(eff)=0)}$. Additionally, we measure the mean $v_{norm}$ of the domain wall on either side of the bubble, since it behaves similarly with response to $B_{y(eff)}$ (Fig. 2). Typical standard uncertainties of $v_{norm}$ after such averaging are about 0.01. Finally, after each bubble measurement, we determine the local $\mu_0 H_c$ of the trilayer by measuring a hysteresis loop at the location of the bubble with a $B_z$ ramp rate of approximately 20 mT·s$^{-1}$. Fig. 3 shows a selection of the $v_{norm}$ curves (gray markers), arranged by $E_{Ar+}$ and $\mu_0 H_c$. We determine $\mu_0 H_{DMI}$ as the negative of the in-plane field which minimizes the quadratic component of $v_{norm}$, defining positive $\mu_0 H_{DMI}$ as acting perpendicular to the surface of the domain wall, in the direction pointing from an $M_z > 0$ region to an $M_z < 0$ region [38,41]. This definition is consistent with positive $\mu_0 H_{DMI}$ leading to right-hand domain walls, and the DMI energy $E_{DMI} = -\mathbf{D} \cdot (\mathbf{S_1} \times \mathbf{S_2})$, with $\mathbf{S_1}$ and $\mathbf{S_2}$ representing the state of adjacent spins, and $\mathbf{D}$ the DMI vector. In Fig. 3, the $\mu_0 H_c \approx 5.0$ mT row exemplifies tuning of $\mu_0 H_{DMI}$ by $E_{Ar+}$.

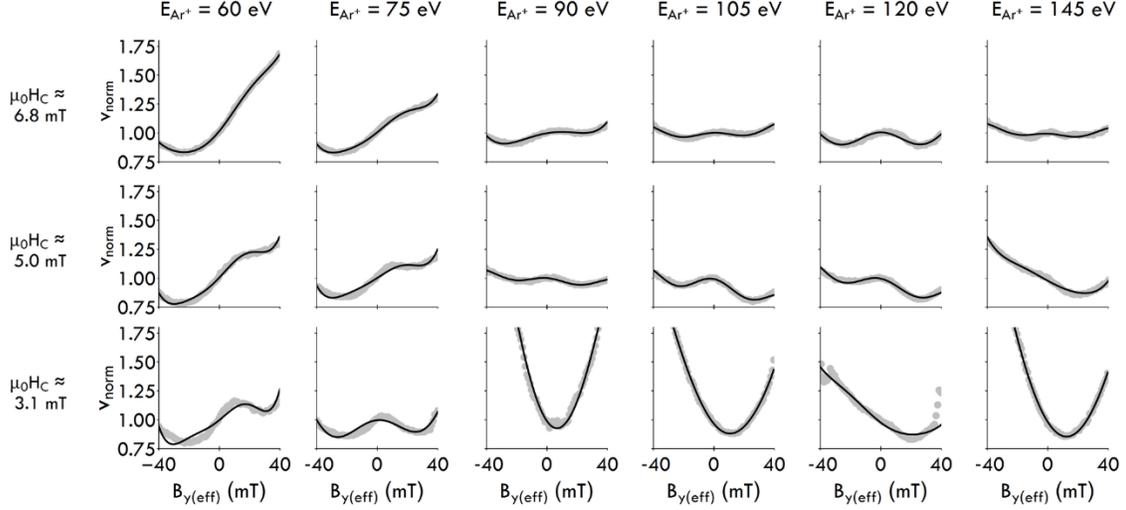

Fig. 3: Representative data shows the normalized domain wall velocity $v_{norm}$ (gray markers) at six values of $E_{Ar+}$ for doses chosen to give three values of $\mu_0H_c$. Black curves are the fits used to extract the value of $K_{pin}$ and $\mu_0H_{DMI}$. The data sets at $\mu_0H_c \approx 5.0$ mT exemplify bipolar tuning of $\mu_0H_{DMI}$ at constant $\mu_0H_c$. Standard uncertainties of $v_{norm}$ are typically 0.01, which is smaller than the data markers in most cases.

Many of the curves show a peak in $v_{norm}$ for $B_{y(eff)} \approx 0$, for example, in the $\mu_0H_c \approx 3.1$ mT, $E_{Ar+} = 75$ eV curve, which prevents fitting this data to a simple quadratic function to accurately extract $\mu_0H_{DMI}$. Previous studies have also reported this peak [31,40-42], ruling out experimental artifacts specific to our experiment.

We propose a physical model as a possible explanation for the peak – a $B_y$-dependent perturbation to the prefactor $v_0$ in the domain wall creep equation, $v = v_0 \exp[-\kappa B_z^{-1/4}]$. In this equation, $\kappa$ is a quantity related to the domain wall surface tension, pinning potential, and temperature. In most cases [38,43,44], changes in $\kappa$ dominate changes in domain wall velocity, but if the exponent is small or $\kappa$ is insensitive to small values of $B_y$, then changes in $v_0$ may dominate [40]. We believe that this is the case here, as the peak amplitude scales differently from the overall parabolic behavior of $v$ in response to a changing $B_z$ (not shown). The value of $v_0$ depends on the attempt frequency $\omega$ of the domain walls in the pinning potential $E_{pin}$ [45,46]. $\omega$ is proportional to the square root of the curvature $\frac{d^2}{d\theta^2}[E_{pin}]$ of the pinning potential. A field such as $B_y$ changes the curvature by moving the configuration of the domain wall away from the minimum of $E_{pin}$. Since the minimum of $E_{pin}$ corresponds to a maximum of $\frac{d^2}{d\theta^2}[E_{pin}]$ for most pinning potentials [40], it also corresponds to a maximum in $\omega$. Therefore, any deviation of the minimum energy configuration of the domain wall from its value at $B_y = 0$ will decrease $\omega$ and thus $v_0$, in an amount proportional to the square root of $\frac{d^2}{d\theta^2}[E_{pin}]$. This explanation predicts a peak in $v_{norm}$ at $B_y = 0$, as is evident if we model the pinning potential by an anisotropy energy minimum with equilibrium angle $\theta_{eq}$ that varies around 0 as the field $B_y$ varies:

$$v_0 \propto \omega \propto \sqrt{\left.\frac{d^2}{d\theta^2}E_{anis}\right|_{\theta=\theta_{eq}}} = \sqrt{2K_{pin}\cos(2\theta_{eq})} = \sqrt{2K_{pin}\cos\left[\frac{2MB_y}{2K_{pin}+MB_z}\right]}. \quad (1)$$

In this model, $M = 5 \times 10^5$ A·m$^{-1}$ is an estimate of the saturation magnetization, which we obtain from a previous study [47], and $K_{pin}$ is the effective anisotropy constant for pinning. Since $v_0$ is outside the exponent in the creep law which predicts domain wall motion, this modification manifests as a multiplicative correction to the domain wall creep law, which is approximately quadratic in $B_y$ at low fields [38]. Therefore, using the method of damped least squares, we fit (Fig. 3, black lines) the $v_{norm}$ curves to the product of a quadratic function which approximates the domain wall creep law with the DMI, and an approximation of relation (1) at $\theta_{eq} \approx 0$ (Supplemental Material S5). This relation fits most of the features in the experimental data, and extracts values of $K_{pin}$ and $\mu_0 H_{DMI}$ that are robust to the details of the fit.

Using this model, we extract a value of $K_{pin}$ = 6 kJ·m$^{-3}$ ± 2 kJ·m$^{-3}$, denoting the mean and standard deviation of 100 $v_{norm}$ curves that show an appreciable central peak. This value is much smaller than typically measured values [43] of the uniaxial anisotropy constant $K_1$, but we can reconcile this discrepancy by noting that the pinning potential should be more complicated than a simple uniaxial anisotropy. We do observe a correlation between $K_{pin}$ and $\mu_0 H_c$ (Supplemental Material S6), supporting our model of the peak in $v_{norm}$ as an effective anisotropy.

The values of $\mu_0 H_{DMI}$ that we extract show systematic variation as a function of $\mu_0 H_c$ and $E_{Ar+}$ (Fig. 4, surface plot), which is the main result of our study – simultaneous control of $\mu_0 H_{DMI}$ and $\mu_0 H_c$. The color scale in Fig. 4 is a linear interpolation between the $\mu_0 H_{DMI}$ values obtained at the $\mu_0 H_c$ and $E_{Ar+}$ conditions that the small black markers indicate. Regions of positive $\mu_0 H_{DMI}$ (blue) and negative $\mu_0 H_{DMI}$ (red) are visible, as is the contour separating them (black line). The values of $\mu_0 H_{DMI}$ have typical standard uncertainties of 1 mT. However, the scatter of the data, manifesting for example as irregularity of the black contour line, indicates that variation in $\mu_0 H_{DMI}$ due to heterogeneity of the trilayer exceeds the measurement uncertainty. We also confirm that Ar$^+$ dose monotonically controls $\mu_0 H_c$ by measuring $\mu_0 H_c$ as a function of Ar$^+$ dose for $E_{Ar+}$ = 100 eV (Fig. 4, inset). Supplemental Material Fig. S7 shows data over the full range of $\mu_0 H_c$ and presents uncertainty evaluation.

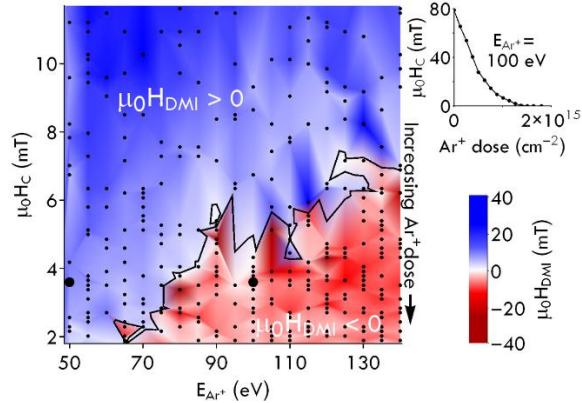

Fig. 4: A surface plot of $\mu_0 H_{DMI}$ shows systematic variation as a function of $\mu_0 H_c$ and $E_{Ar+}$. Ar$^+$ dose increases toward the bottom of this surface plot. The black contour indicates the boundary between positive (blue) and negative (red) $\mu_0 H_{DMI}$. Circular black markers indicate the $\mu_0 H_c$ and $E_{Ar+}$ values of the measurements between which the color map interpolates. The two large black markers indicate the approximate parameters corresponding to Supplemental Material S8. Inset: $\mu_0 H_c$ decreases monotonically with Ar$^+$ dose for a sample with $E_{Ar+}$ = 100 eV.

To demonstrate spatial resolution of this tuning, we prepare two nearby regions of the trilayer to have the same $\mu_0 H_c$ of 3.6 mT, but different values of $\mu_0 H_{DMI}$, 20 mT and -20 mT, with exposure to 50 eV

Ar$^+$ and 100 eV Ar$^+$, respectively (large black markers, Fig. 4). These regions are close enough such that the effects of positive and negative $\mu_0H_{DMI}$ are evident with identical $B_y$ and $B_z$ excitation, in the same field of view of the MOKE microscope (Supplemental Material S8).

While both $\mu_0H_{DMI}$ and $\mu_0H_c$ depend on spin-orbit coupling at the Pt/Co interfaces, our results indicate that they depend differently on the details of the layers and interfaces. A stopping and range of ions in matter (SRIM) simulation [48] (Supplementary Material S9) shows that around $E_{Ar+}$ = 100 eV, Ar$^+$ primarily influences the top Pt and Co layers. Auger spectroscopy (Supplementary Material S9) does not show evidence of Ar implantation, and indicates that the change of the sign of $\mu_0H_{DMI}$ is associated with removal of the top Pt monolayer. Although the different behavior of $\mu_0H_{DMI}$ and $\mu_0H_c$ with $E_{Ar+}$ and Ar$^+$ dose may be complicated, previous studies provide some guidance. Several studies [33-36] have shown that ion-induced interfacial disorder, and reduction of the thickness of the top Pt layer [49], decrease $\mu_0H_c$, while $\mu_0H_{DMI}$ is primarily sensitive to the interfacial Pt monolayers [53]. Therefore, removal of the top Pt could affect $\mu_0H_{DMI}$ in two ways, eliminating the positive influence of the layer on $\mu_0H_{DMI}$ [50], and allowing oxidation of the Co [51,52].

Eliminating the influence of the top Pt should have a strong influence on $\mu_0H_{DMI}$. The etch rate of Pt vanishes at lower $E_{Ar+}$ [54], and depends differently on $E_{Ar+}$ than interfacial disorder does. This suggests that, for a given thickness of etched Pt, higher energy Ar$^+$ causes less interfacial disorder than lower energy Ar$^+$, causing an $E_{Ar+}$ dependence of $\mu_0H_{DMI}$ at fixed $\mu_0H_c$. For example, exposure to 130 eV Ar$^+$ leads to a large negative $\mu_0H_{DMI}$ after removal of the top Pt layer, but 50 eV Ar$^+$ causes enough disorder to drive the film to the spin reorientation transition at doses necessary to remove the top Pt layer.

To test the effect of Co oxidation on $\mu_0H_{DMI}$, we deposit a layer of Au with a thickness of approximately 10 nm immediately after Ar$^+$ irradiation in vacuum. After exposure to air, we measure similar values of $\mu_0H_{DMI}$ with and without the Au layer (Supplementary Material S10). This suggests that Pt removal is the primary mechanism for our simultaneous control, but does not rule out oxidation as a factor, as we are not certain that the Au layer is hermetic, and we do not know of any influence of the Au itself on $\mu_0H_{DMI}$. Measurements in vacuum could elucidate these effects.

In conclusion, we have demonstrated the use of Ar$^+$ irradiation around 100 eV to tune the sign and magnitude of the DMI in ultrathin Pt/Co/Pt trilayers. The tuning is spatially variable, separately from $\mu_0H_c$. At low values of the DMI, we observe a peak in domain wall velocity which we propose to explain by a modification of the depinning attempt frequency in the creep regime of domain wall motion. This model may also explain a discrepancy between measurements of $\mu_0H_{DMI}$ by domain wall motion and Brillouin light scattering [41], which do not account for the influence of the peak in domain wall velocity. Our technique for simultaneous control of $\mu_0H_{DMI}$ and $\mu_0H_c$ on a single chip enables systematic study of the effects of the DMI in isolation from stronger interactions [55], and potentially allows micro- and nanopatterning of the DMI [56]. Finally, this new level of control will enable magnetic materials with engineered DMI for proposed and existing technological applications [57-59].

**Acknowledgments**
The authors acknowledge Edward Cazalas for performing and interpreting SRIM simulations, Kerry Siebein for assistance with X-ray diffraction measurements, and Carl Boone, Paul Haney, Daniel Gopman, and J. Alexander Liddle for thoughtful reviews and helpful criticism. K.W.K acknowledges Hyun-Woo Lee for fruitful discussion. A.L.B. and K.W.K. acknowledge support of this research under the Cooperative Research Agreement between the University of Maryland and the National Institute of Standards and Technology Center for Nanoscale Science and Technology, Award No. 70NANB10H193, through the University of Maryland. K.W.K also acknowledges support by Basic Science Research Program through the National Research Foundation of Korea (NRF) funded by the Ministry of Education (2016R1A6A3A03008831).

**Supplemental material** *for* **Simultaneous control of the Dzyaloshinskii-Moriya interaction and magnetic anisotropy in nanomagnetic trilayers**

A.L. Balk[1,2,3], K-W. Kim[1,2], D.T. Pierce[1], M.D. Stiles[1], J. Unguris[1], S.M. Stavis[1],*

[1]Center for Nanoscale Science and Technology, National Institute of Standards and Technology, Gaithersburg, MD 20899 USA, [2]Maryland NanoCenter, University of Maryland, College Park, MD 20742 USA, [3]National High Magnetic Field Laboratory, Los Alamos National Laboratory, Los Alamos, NM 87545 USA, *Address correspondence to sstavis@nist.gov

**Index**
S1: Sample preparation
S2: Field excitation
S3: Real-time image processing
S4: Absence of chiral damping
S5: Fit function
S6: Correlation of pinning anisotropy and coercive field
S7: Measurement of Dzyaloshinskii-Moriya interaction (DMI) over full range of coercive field
S8: Simultaneous observation of opposite signs of DMI on the same chip
S9: Effects of Ar[+] irradiation
S10: Au layer
References

**S1: Sample preparation**

We sputter the Pt/Co/Pt trilayer on a p-type silicon substrate by the following process. First, we perform *in situ* Ar[+] etching at a power of 50 W for a duration of 120 s to remove any contaminants from the substrate. Next, we grow the bottom Pt layer with a base pressure of $5 \times 10^{-4}$ Pa, a sputtering pressure of 6.6 Pa, and a power of 100 W, which is optimal for film thickness uniformity. Then, we grow the Co layer at a sputtering pressure of 6.6 Pa and a power of 65 W. Finally, we grow the top Pt layer at a sputtering pressure of 6.6 Pa and a power of 60 W. The lower power for growth of the Co and top Pt layers reduces physical damage from atomic and ionic impacts to the existing interfaces of the layers [1]. All sputtering steps use a DC Ar[+] plasma at a flow rate of 0.83 atm·cm$^3$·s$^{-1}$.

We determine the thickness of the top Pt and Co layers by growing test films under identical sputtering conditions but longer deposition times, measuring their thickness by atomic force microscopy, and calculating their growth rates. In this way, we determine that the top Pt layer has a growth rate of 0.09 nm·s$^{-1}$ ± 0.01 nm·s$^{-1}$ and a thickness of 1.7 nm ± 0.3 nm, and the Co layer has a growth rate of 0.13 nm·s$^{-1}$ ± 0.02 nm·s$^{-1}$ and a thickness of 0.8 nm ± 0.1 nm. We propagate the standard uncertainty from test film thickness to top Pt and Co layer thickness. We measure X-ray reflectivity spectra of the trilayer (not shown) and use commercial software to fit them to a model that iteratively determines the thickness of the bottom Pt film as 35 nm ± 2 nm, which is consistent with the growth rate of the top Pt layer. We derive this limit of uncertainty from the uncertainties of the trilayer properties that we input into the model, such as interface roughness and thickness of the other layers.

The $\mu_0 H_C$ of the trilayer as grown is approximately 80 mT, which is larger than the maximum field of the electromagnet that we use to apply $B_z$. Therefore, we do not measure the DMI in the trilayer as grown. After trilayer growth, we dice this wafer into chips to investigate the effects of Ar[+] irradiation.

**S2: Field excitation**

We use a bipolar, pulsed $B_z$ waveform at a repetition frequency of 50 Hz to 150 Hz and an amplitude of 1 mT to 20 mT to measure magnetic bubble domains and the effective Dzyaloshinskii-Moriya interaction (DMI) field $\mu_0 H_{DMI}$ (Fig. S2). Each cycle of the $B_z$ waveform starts with a negative field pulse which saturates the magnetic state of the trilayer. Immediately after, a smaller positive pulse nucleates and expands bubbles for measurement. Finally, the magnetic field returns to a small, negative constant field, and remains at that field for bubble observation until the next pulse. The constant negative field ensures complete initialization after each measurement cycle, but is small enough that the domain walls under observation are approximately stationary. The duration of the constant field portion of the waveform is much larger than the total duration of the positive and negative pulses, so the image contrast is close to that of images of stationary bubbles. This $B_z$ waveform induces bubbles at the repetition frequency, which is much higher than the imaging frequency of 10 Hz of our charge-coupled device (CCD) camera. In this way, each MOKE micrograph is the average of 5 to 15 bubble expansions, reducing noise by averaging over stochastic domain wall pinning. This $B_z$ waveform allows continuous observation of bubble growth at tens of hertz.

During excitation with this $B_z$ waveform, each chip shows nucleation sites at an areal density of approximately $10^2$ mm$^{-2}$, enabling measurement of isolated bubbles. Earlier studies [2] have used more complicated protocols for $B_z$ excitation, based on a nucleation pulse followed by measurement during a propagation pulse to eliminate possible effects due to inconsistent nucleation of bubbles. Here, we use only a single pulse to make the measurement faster. We verify that the results of single and double pulses are consistent (not shown).

For excitation in the plane of the trilayer, we apply $B_y$ in a triangular waveform at a frequency of 50 mHz and an amplitude of 40 mT. This $B_y$ frequency is much lower than the $B_z$ repetition frequency and the imaging frequency, so that the position of each domain wall accurately indicates the influence of $B_y$ on domain propagation velocity. The bubbles are sensitive to $z$-direction misalignment of $B_y$, so we adjust the angle of the trilayer at large values of $B_y$ to minimize variations in the size of the expanded bubbles. We can measure single bubbles consistently for $B_y$ up to approximately 20 mT to 40 mT. Higher values result in nucleation sites that are overly dense for measurement of single bubbles.

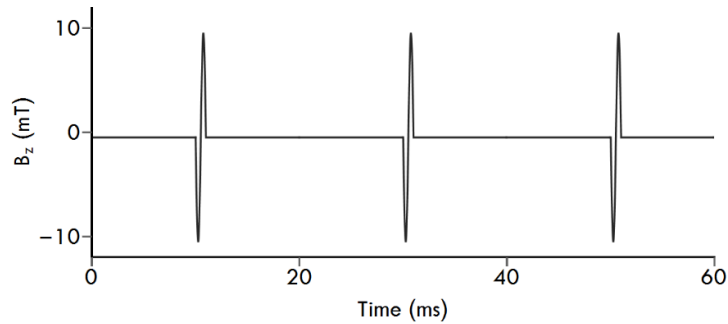

Fig. S2: A simulation of the $B_z$ excitation scheme shows a series of bipolar pulses that first saturate the sample in the negative direction and then drive domain walls for the measurement. The domain walls are approximately stationary during the long constant field portions of the waveform, so the contrast of the video data approaches that of still images of stationary bubbles.

**S3: Real-time image processing**

Bubbles appear in unprocessed MOKE micrographs as roughly circular patches (Fig. S3). Each MOKE micrograph averages over 5 to 15 bubble expansions, due to the repetition frequency of the $B_z$ waveform of 50 Hz to 150 Hz and the imaging frequency of the CCD camera of 10 Hz. In real time, we extract rectangular sections from these micrographs (Fig. S3, dashed line), average these sections in the vertical direction to generate one-dimensional profiles of the bubbles, and fit these profiles with a sum of two offset arctangent functions that model the domain walls. This empirical fit is a good approximation of the effects of stochastic domain wall pinning and broadening by optical diffraction, and provides super-resolution measurements of domain wall positions. This fit includes only the two domain wall positions as floating parameters to increase measurement speed, enabling the measurement of approximately 400 bubbles for this study and mapping of the DMI. The fit provides domain wall displacements with standard uncertainties of approximately 400 nm at 10 Hz. We calculate domain wall velocities by dividing the domain wall displacement by the $B_z$ pulse duration, resulting in values of 10 mm·s$^{-1}$ to 50 mm·s$^{-1}$, with typical standard uncertainties of 0.8 mm·s$^{-1}$.

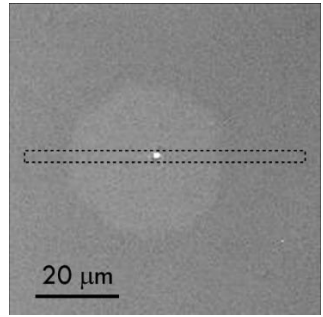

Fig S3: An unprocessed MOKE micrograph shows a magnetic bubble domain. In real time, we extract rectangular sections from these micrographs, outlined above with a dashed black line, average these sections in the vertical direction to generate one-dimensional profiles of the bubbles, and fit the profiles with a sum of two offset arctangent functions that empirically model the domain walls.

**S4: Absence of chiral damping**

A recent study [3] proposed that damping of spin precession, depending on the chirality of the spin texture and separately from the effects of DMI, could lead to a non-zero minimum in domain wall velocity as a function of in-plane field. To investigate this possibility, we replot the domain wall velocity curves from Fig. 2 as a function of $B_y$, with one curve offset along the x-axis (Fig. S4). If chiral damping were a factor in these measurements, then there would be a significant asymmetry between the two curves, which we do not observe.

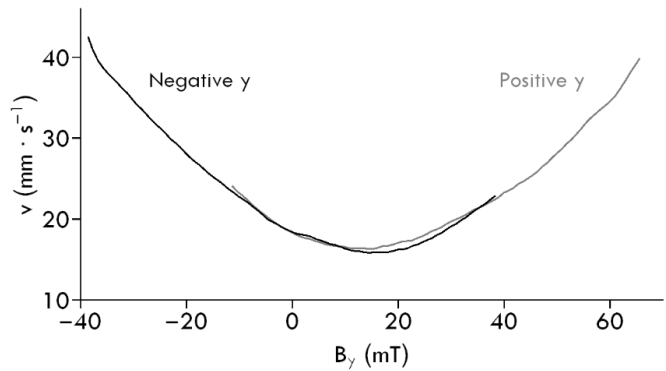

Fig. S4: A damping term which depends on the chirality of the domain wall would manifest as a significant asymmetry between the negative y domain wall velocity curve (black) and the positive *y* domain wall velocity curve which we have offset along the x-axis (gray). We do not observe this asymmetry in our data, ruling out the effects of chiral damping in our samples. Standard uncertainties for the data in Fig. S4 and Fig. 2 are approximately 0.2 mm·s$^{-1}$, which we determine from the standard deviations of the values of each resampled point.

**S5: Fit function**

We assume that the equilibrium angle between the pinned moment within the domain wall and the surface normal,

$$\theta_{eq} \approx \frac{2MB_y}{2K_{pin} + MB_z},$$

is close to zero. For numerical convenience in the fitting procedure we approximate $v_0$ around $\theta_{eq} = 0$,

$$v_0 \propto \sqrt{2K_{pin} \cos\left[\frac{2MB_y}{2K_{pin} + MB_z}\right]} = \sqrt{2K_{pin}} \cdot \sqrt{(1-\alpha)} \approx \sqrt{2K_{pin}} \cdot (1-\frac{\alpha}{2}) = \sqrt{\frac{K_{pin}}{2}} \cdot \left[1 + \cos\left[\frac{2MB_y}{2K_{pin} + MB_z}\right]\right],$$

where

$$\alpha = 1 - \cos\left[\frac{2MB_y}{2K_{pin} + MB_z}\right].$$

We fit our experimental $v_{norm}$ curves with this expression, absorbing the prefactor and proportionality constant into a single fit parameter representing the amplitude of the peak.

**S6: Correlation of pinning anisotropy and coercive field**

We extract a value of the anisotropy constant for pinning $K_{pin}$ from each domain wall velocity $\tilde{v}$ curve that shows a peak near $B_{y(eff)} = 0$. We find a correlation between values of $K_{pin}$ and local values of $\mu_0 H_c$, with a Pearson's correlation coefficient (PCC) of approximately 0.3 (Fig. S6).

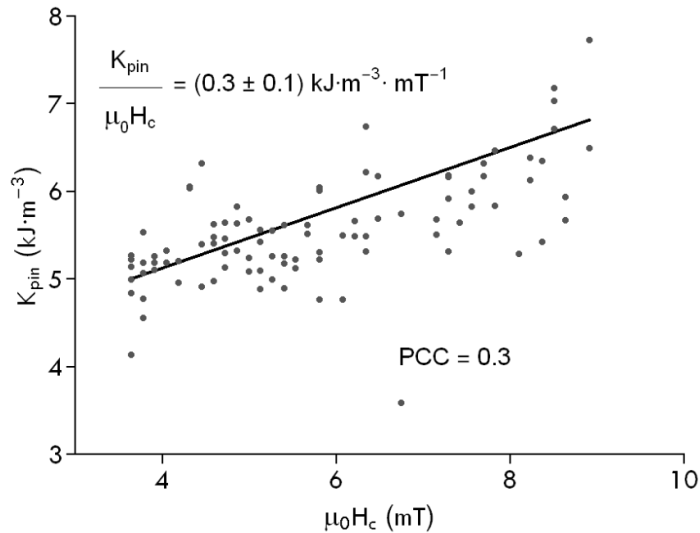

Fig. S6: Values of $K_{pin}$ from domain wall velocity curves show a correlation with locally measured values of $\mu_0 H_c$. Standard uncertainties for $K_{pin}$ values are around 1 kJ·m$^{-3}$, which we determine from the fits.

**S7: Measurement of Dzyaloshinskii-Moriya interaction (DMI) over full range of coercive field**

A surface plot (Fig. S7) of the DMI field $\mu_0 H_{DMI}$ over the full range of data from our study. The full range of data shows that positive $\mu_0 H_{DMI}$ persists to high coercive field $\mu_0 H_c$.

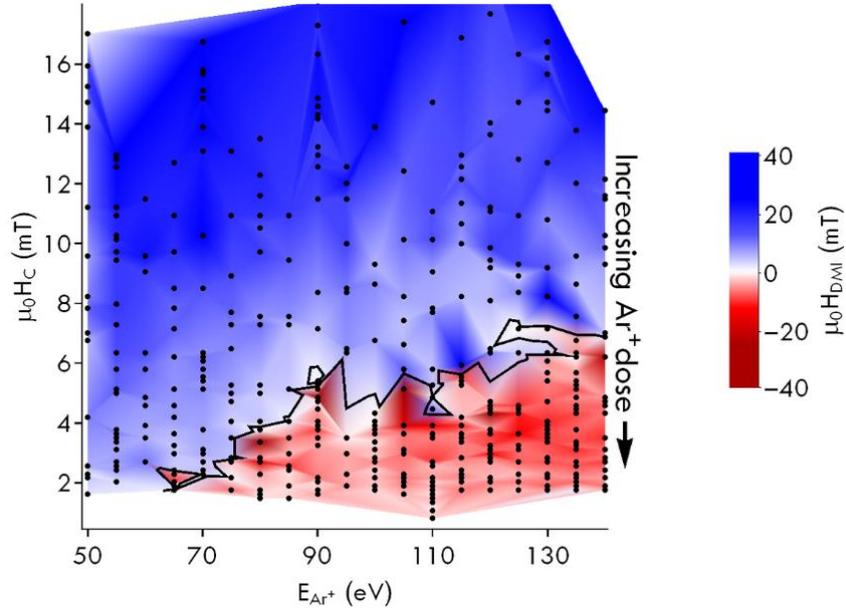

Fig. S7: A surface plot shows the full range of $\mu_0H_{DMI}$ values that we measure as a function of $\mu_0H_c$ and $E_{Ar+}$. The black contour indicates the interpolated boundary between positive (blue) and negative (red) $\mu_0H_{DMI}$. Black markers are measurement results which form the basis of the color map, which is a linear interpolation. Standard uncertanties of $\mu_0H_{DMI}$ have a mean value of approximately 1 mT, which we derive from the fits for the data in this plot. The limit of uncertainty of the $\mu_0H_c$ values is 0.3 mT, which we determine by the field sampling rate when performing the hysteresis measurements on the film. The limit of uncertainty of the $E_{Ar+}$ values is 1 eV, which we estimate from the uncertainty of the energy adjustment mechanism of the Ar+ source.

**S8: Simultaneous observation of opposite signs of DMI on the same chip**

Supplemental S8 shows bubble growth in two regions of the same chip with opposite signs of $\mu_0H_{DMI}$. We expose the region on the left to 50 eV Ar+ and the region on the right to 100 eV Ar+ (Fig. S8, top). We prepare the two regions to have similar $\mu_0H_c$, enabling simultaneous observation with the same $B_z$ excitation waveform. In Fig. S8, the $B_z$ waveform is as we describe in Supplementary S2, such that the growth of the bubbles is invisible but the relative velocity of the domain walls during growth is detectable by their positions. Magnified regions of interest within the yellow boxes are at the bottom of Fig. S8. These show bubbles under exposure to an in-plane field $B_y$ with an amplitude of approximately 40 mT in two different directions. The opposite sign of $\mu_0H_{DMI}$ for the two regions is evident as the opposite direction of bubble expansion in response to the same $B_y$. Fig. S8 has opposite contrast relative to the other data in the rest of the manuscript, so the bubbles appear as dark on a light background.

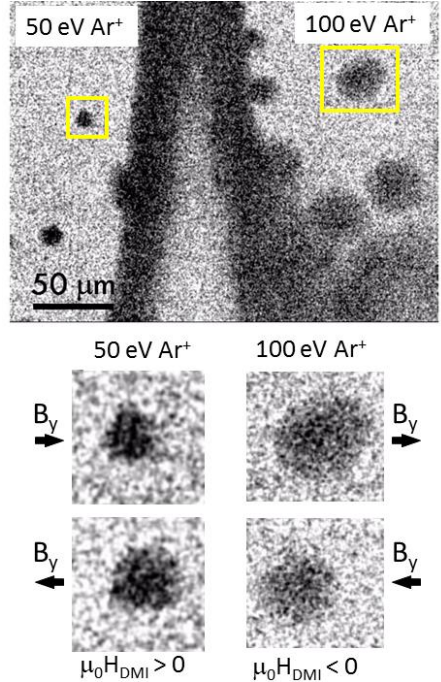

Fig. S8: Two signs of $\mu_0 H_{DMI}$ on the same chip of the Pt/Co/Pt trilayer. Top: A MOKE micrograph shows bubbles in two regions of the trilayer. Yellow boxes highlight two bubbles which demonstrate the opposite sign of the DMI. Bottom: Magnified images of these bubbles in response to a positive direction and negative direction $B_y$ with approximate magnitude of 40 mT. The opposite sign of the DMI is evident as opposite direction of bubble movement in response to the same $B_y$ and $B_z$.

**S9: Effects of Ar⁺ irradiation**

We perform a stopping and range of ions (SRIM) simulation of Ar⁺ penetration into the trilayer. The simulation is of 100 eV Ar⁺ incident normally on a Pt/Co(0.8 nm)/Pt(0.3 nm) trilayer with a Pt density of 21 g·cm⁻³ and a Co density of 8.9 g·cm⁻³. The reduction in thickness of the top Pt from the as-grown value of 1.7 nm accounts for its removal by an experimental Ar⁺ dose of approximately $1 \times 10^{15}$ cm⁻². This simulation shows that approximately 97 % of Ar⁺ ions stop before the bottom Co/Pt interface (Figure S9, top), indicating that the Ar⁺ irradiation has a larger influence on the top Pt/Co interface.

We use Auger spectroscopy to determine the elemental composition of the top surface of the trilayer. We take Auger spectra from areas of the trilayer irradiated with different doses of 100 eV Ar⁺, without exposing the trilayer to air. At doses larger than approximately $2 \times 10^{14}$ cm⁻², the Auger spectra exhibit three peaks at energies characteristic of Co (Fig. S9, middle, mean of all spectra taken in this study), but they do not show evidence for Ar implantation in our trilayers (Fig. S9, middle, inset). This is consistent with the chemically inert Ar diffusing out of the trilayer after irradiation. The integrated intensity of the Co peaks is a qualitative indicator of the fraction of Co on the surface of the sample, which increases until a dose of approximately $20 \times 10^{14}$ cm⁻² (Fig. S9, bottom), indicating that this dose fully etches the top Pt layer. This is consistent with the etch rate of Pt in response to 100 eV Ar⁺ in our system, as well as the thickness of the top Pt layer. We then characterize perpendicular anisotropy and the DMI on the same areas to correlate surface composition with the DMI. The $\mu_0 H_{DMI} = 0$ point occurs at doses less than those that cause the Co peaks to have maximum intensity, indicating that the $\mu_0 H_{DMI} = 0$ point occurs after the removal of some, but not all, of the top Pt monolayer. This plot also indicates the dose of Ar⁺ exposure at this energy to reach the spin reorientation transition (SRT) of the trilayer.

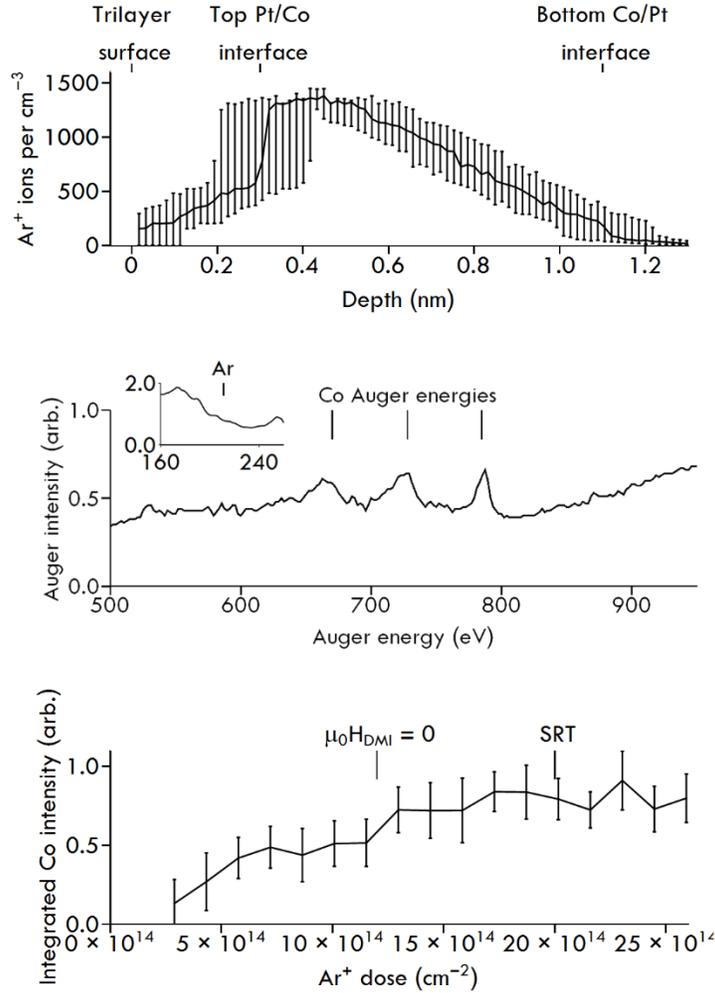

Fig. S9: Effects of Ar$^+$ irradiation on the trilayer. Top: SRIM simulation results indicate that approximately 97 % of 100 eV Ar$^+$ ions do not penetrate to the bottom Co/Pt interface, therefore influencing the top Pt/Co interface more. Vertical bars are limits of uncertainty corresponding to the unknown thickness of the top Pt layer. Middle: Three peaks in the Auger energy spectrum indicate the presence of Co in the top 1 nm of the trilayer. This spectrum is the mean of all spectra from this Auger study. Inset: Auger spectra do not show evidence of Ar implantation, which would manifest as a peak at 211 eV. The axes of the inset are the same as those of the main plot. Typical values of measurement uncertainty for Auger intensity are approximately 0.01, in the arbitrary units of the plot. We estimate this uncertainty as the standard deviation of Auger intensities at regions of the spectra with no evident peaks. Bottom: The integrated intensity of the three peaks increases above the background at 100 eV Ar$^+$ doses of larger than approximately $2 \times 10^{14}$ cm$^{-2}$, and stops increasing after a dose of approximately $20 \times 10^{14}$ cm$^{-2}$. The increasing integrated intensity of the Co peaks indicates the removal of the top Pt layer, and coincides with $\mu_0 H_{DMI} = 0$, which we measure after Auger spectroscopy with MOKE microscopy. The tick mark labeled "SRT" indicates the Ar$^+$ dose required to reach the spin reorientation transition. Vertical bars are standard uncertainties, which we obtain from the spectral values integrated to obtain the data.

**S10: Au layer**

We investigate the possible effects of Co oxidation on $\mu_0 H_{DMI}$ in a trilayer region with a nominal $\mu_0 H_C$ of 2 mT and $E_{Ar+}$ = 100 mT. Auger spectroscopy (S9) shows that, with these parameters, the trilayer has a partially exposed Co top surface and an opposite sign of $\mu_0 H_{DMI}$ relative to regions of the trilayer with an intact Pt top surface, so any effects of oxidation should be present. We deposit a layer of Au with a nominal thickness of 10 nm on part of this region after Ar$^+$ irradiation, without breaking vacuum and exposing the sample to air. This Au layer is thick enough to reduce oxidation of Co but thin enough to allow light transmission for MOKE microscopy, and has a visible optical density. After exposure to air, we measure three bubbles on a region with the Au layer, and three bubbles on a region without the Au layer, averaging the $v_{norm}$ curves (Fig. S10). We find that $\mu_0 H_{DMI}$ is similar for these two regions, suggesting that oxidation does not play a dominant role in altering $\mu_0 H_{DMI}$.

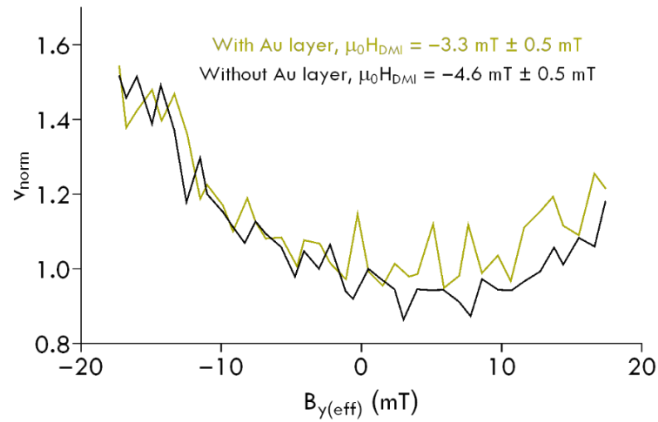

Fig. S10: Comparison of $v_{norm}$ curves from a region of the trilayer with (gold) and without (black) an Au layer to reduce Co oxidation. The $\mu_0 H_{DMI}$ values from fitting these curves, -3.3 mT ± 0.5 mT and -4.6 ± 0.5 mT are similar. We derive standard uncertainties of $v_{norm}$ from the fits to extract $v_{norm}$, and standard uncertainties of $\mu_0 H_{DMI}$ by propagating the uncertainty of the $v_{norm}$ measurement through the quadratic fit to extract $\mu_0 H_{DMI}$.